\documentclass[prd,twocolumn,showpacs,preprintnumbers,nofootinbib,amsmath,amssymb,superscriptaddress,nobalancelastpage]{revtex4-1}

\usepackage{graphicx}
\usepackage{amsmath}
\usepackage{amssymb}

\font\FermiSmallfont=cmssq8 scaled 1200
\def\LANLppthead#1#2{
\null 
\begin{center}\vskip -1.0truein{\hbox to 7.5truein {
\hfill
\vbox to 1in {\vfill \FermiSmallfont
              \hbox{#1}
              \hbox{#2}
              \vfill}
}}\vskip-0.0truein\end{center}}

\def\Msun{M_\odot}

\def\rhomax{\rho_{\rm max}}

\def\VEV#1{\langle #1 \rangle}
\def\vsat{v_{\rm sat}}

\begin{document}

\title{Galactic Substructure and Dark Matter Annihilation in the
       Milky Way Halo}
\author{Marc Kamionkowski}
\affiliation{California Institute of Technology, Mail Code
350-17, Pasadena, CA 91125}\email{kamion@tapir.caltech.edu}
\author{Savvas M. Koushiappas}
\affiliation{Department of Physics, Brown University, 182 Hope
Street, Providence, RI 02912} \email{koushiappas@brown.edu}
\author{Michael Kuhlen}
\affiliation{Theoretical Astrophysics Center, University of California Berkeley, Berkeley, CA 94720} \email{mqk@astro.berkeley.edu}

\begin{abstract}
We study the effects of substructure on the rate of dark-matter
annihilation in the Galactic halo. We use an analytic model for
substructure that can extend numerical simulation results to scales
too small to be resolved by the simulations.  We first calibrate the
analytic model to numerical simulations, and then determine the
annihilation boost factor, for standard WIMP models as well as those
with Sommerfeld (or other) enhancements, as a function of
Galactocentric radius in the Milky Way. We provide an estimate of the
dependence of the gamma-ray intensity of WIMP annihilation as a
function of angular distance from the Galactic center.  This
methodology, coupled with future numerical simulation results can be a
powerful tool that can be used to constrain WIMP properties using {\em
  Fermi} all-sky data.
\end{abstract}

\maketitle

\section{Introduction}

Weakly-interacting massive particles (WIMPs) provide perhaps the most
promising class of dark-matter candidates.  These are particles that
arise in theories of new electroweak-scale physics, such as low-energy
supersymmetry (SUSY) \cite{Jungman:1995df} or models with universal
extra dimensions (UEDs) \cite{Hooper:2007qk}.  Experiments that seek
to directly detect these particles in low-background experiments, or
to indirectly detect them through observation of energetic neutrinos
from WIMP annihilation in the Sun, are now beginning to dig into the
favored WIMP parameter space.  However, there are also prospects for
indirectly detecting WIMPs through observation of gamma rays and/or
cosmic-ray positrons, antiprotons, or antideuterons from WIMP
annihilation in the Galactic halo.  These annihilation products have
received considerable attention in the recent literature with the
attribution of some reported (and still controversial) cosmic-ray
anomalies
\cite{Finkbeiner:2003im,*Dobler:2007wv,*Hooper:2007kb,*Jean:2003ci,*Knodlseder:2003sv,*Weidenspointner2004,*Strong:2005zx,*Thompson2005,*Adriani:2008zr}
to WIMP-annihilation products.

The total rate at which WIMPs annihilate in the Galactic halo is
proportional to the volume integral of the square of the
dark-matter density.  In the canonical (and simplest) model, the
dark matter is smoothly distributed in the halo in a
spherically-symmetric way with a dark-matter density $\rho(r)$
that is a monotonically decreasing function of $r$; for example,
it is often modeled as an isothermal or a
Navarro-Frenk-White (NFW) profile \cite{Navarro:1996gj}.

However, analytic arguments and numerical simulations show that there
should be substructure in the dark-matter distribution in the Galactic
halo
\cite{Klypin:1997fb,*Ghigna:1998vn,*Klypin:1999uc,*Moore:1999nt,*Boehm:2000gq,*Green:2003un,*Green:2005fa,*Green:2005kf,*Bertschinger:2006nq,*Berezinsky:2007qu,*Diemand:2007qr,Loeb:2005pm,Diemand:2005vz,*Gao:2005hn,*Diemand:2006ey}.
The dark matter may be clumped; some of it may be bound in
higher-density self-bound subhalos; and some may be in tidal streams
\cite{Stiff:2001dq,*Freese:2003tt,*Zemp:2008gw}.  This is an outcome
of hierarchical clustering, in which small dense halos form first and
then merge to form progressively larger structures.  If dark matter in
the halo is clumped, then the total annihilation rate will be enhanced
by some boost factor
\cite{Bergstrom:1998jj,*Bergstrom:1998zs,*CalcaneoRoldan:2000yt,*Tasitsiomi:2002vh,*Berezinsky:2003vn,*Tasitsiomi:2003ue,*Stoehr:2003hf,*Koushiappas:2003bn,*Baltz:2006sv,*Pieri:2005pg,*Koushiappas:2006qq,*Diemand:2006ik,*Berezinsky:2006qm,*Pieri:2007ir,*Kuhlen:2008aw,*Robertson2009,*Kuhlen2009kx},
the increased rate per unit volume in dense regions outweighing the
decreased rate per unit volume in lower-density regions.

In the canonical WIMP scenario, the substructure may have a roughly
scale-invariant distribution in substructure mass/size extending all
the way down to substructures on mass scales $\sim10^{-10}~M_\odot$
\cite{Chen:2001jz,Profumo:2006bv,Diemand:2005vz,*Gao:2005hn,*Diemand:2006ey},
roughly 22 orders of magnitude smaller than the $\sim10^{12}~M_\odot$
Milky Way halo.  While state-of-the-art numerical calculations now
have the resolution to simulate several decades in this hierarchy,
they are very far from being able to follow the survival and evolution
of the smallest substructures through all of the generations in the
structure-formation hierarchy that result in a Milky Way halo.
Analytic calculations of these survival probabilities are difficult
\cite{Zhao:2005mb,*Zhao:2005py,*Berezinsky:2005py,Goerdt:2006hp}.

However, given that the smallest subhalos are likely to be the
densest, the boost factor may depend significantly on the existence of
these substructures.  This is particularly true in models with a
Sommerfeld enhancement in the annihilation rate
\cite{Hisano:2004ds,*MarchRussell:2008yu,*Feng:2008mu,*Feng:2008ya,*ArkaniHamed:2008qn,*Pospelov:2008jd}---those
where the annihilation rate increases with lower WIMP
velocities---since the smallest subhalos are also likely to have the
smallest velocity dispersion.

In a previous paper \cite{Kamionkowski:2008vw}, we presented an
analytic model to describe the self-similar substructure expected from
hierarchical clustering.  The model predicted a high-density power-law
tail for the probability for a given point in the halo to be in a
clump of density $\rho$.  Subsequent to that paper, another appeared
\cite{Vogelsberger2009} presented N-body simulations that showed this
power-law tail.  After calibration to numerical simulations, the model
can be used to extrapolate the results of numerical simulations to
substructure-mass scales far smaller than those currently accessible
to the simulations.  We investigated in
Ref.~\cite{Kamionkowski:2008vw} the dependence of the boost factor
assuming a canonical WIMP-annihilation rate (i.e., no Sommerfeld
enhancement). A complementary analytical approach to this problem,
based on the stable clustering hypothesis, was recently presented by
\cite{Afshordi2009}.

In this paper we re-visit and extend those calculations.  We
first describe in Section \ref{sec:model} the analytic model.
We extend the model by including a finite width for the smoothly
distributed component of dark matter in addition to the
high-density power-law tail.  We then use in Section
\ref{sec:calibration} state-of-the-art
N-body simulations to measure the width of the smooth component
and the amplitude of the power-law tail and thus calibrate the
analytic model to current simulations.
We moreover determine how the substructure distribution
varies with Galactocentric radius in the halo.  In Section
\ref{sec:boostfactor} we use
the calibrated model to determine the boost factors for canonical WIMPs
and for WIMPs with a Sommerfeld enhancement to the annihilation
rate.  Section \ref{sec:wimpmodels} reviews the
substructure-model parameters expected from WIMP models, and
Section \ref{sec:angular} determines the angular dependence of
the intensity of gamma-ray radiation both with and without the
substructure boost factor we obtain.  Section
\ref{sec:discussion} reviews the model and results and then
provides some comments, caveats, and directions for future
development of the model.

\section{Substructure Model}
\label{sec:model}

Before discussing our analytic model for substructure we begin 
by presenting the canonical smooth-halo model against
which the substructure model will be compared.  We take as the
canonical model for the Galactic halo an NFW profile,
\begin{equation}
     \bar\rho(r) = \frac{4 \; \rho_s}{(r/r_s)( 1+ r/r_s)^2},
\label{eq:NFW}
\end{equation}
as a function of Galactocentric radius $r$, with parameters
$\rho_s=0.051$~GeV~cm$^{-3}$ and $r_s=21.7$~kpc taken to provide a
reasonable fit to the Milky Way rotation curve.

If dark matter in the halo is clumped, then the densities at all
points with the same $r$ will not necessarily be the same.
Instead, there will be some probability distribution function
$P(\rho,r)$ defined so that $P(\rho,r)d\rho$ is the probability
that a particular point in the Galactic halo (at some fixed
Galactocentric radius $r$) has a density between $\rho$ and
$\rho+d\rho$.  According to the arguments of
Ref.~\cite{Kamionkowski:2008vw}, a fraction $f_s$ of the volume
of the halo (in fact most of the volume; $f_s\simeq1$, as we
will see) should be filled with a smooth dark-matter component
with density $\rho_h$, and a fraction $1-f_s \ll 1$ will consist
of a high-density clumped component, with something like a
power-law distribution of densities.  Thus, the probability
distribution function that we use is
\begin{eqnarray}
     P(\rho;r) =& & \frac{f_s}{\sqrt{2\pi \, \Delta^2}} \, \frac{1}{\rho}
     \, \exp \left\{ -\frac{1}{2 \Delta^2} \left[
     \ln\left(\frac{\rho}{\rho_h} e^{\Delta^2/2 }\right)\right]^2 \right\}
      \nonumber \\
     + & &\left(1-f_s\right)
     \frac{1+\alpha(r)}{\rho_h}  \Theta\left(\rho-\rho_h\right)
     \left( \frac{\rho}{\rho_h} \right)^{-(2+\alpha)},
\label{eqn:pdf}
\end{eqnarray}
where $\Theta(x)$ is the Heaviside step function.  Here, the first
term describes the smooth host halo component as having a Gaussian
distribution in $\ln \rho$ with mean density $\rho_h$ and a Gaussian
width in $\ln\rho$ of $\Delta$.  This distribution peaks at a density
$\rho_0=e^{-3\Delta^2/2} \rho_h$, a density slightly below
$\rho_h$.

The second term in Eq.~(\ref{eqn:pdf}) is
the high-density power-law tail due to substructures that remain
from earlier generations in the structure-formation hierarchy.  The
parameter $\rho_h$ in Eq.~(\ref{eqn:pdf}) is in fact a function
of $r$; we will see below that $\rho_h(r)\simeq \bar\rho$ to
very good accuracy.  There may also be further $r$ dependence in
$P(\rho;r)$ through an $r$ dependence of the parameters
$\Delta$, $\alpha$, and $f_s$.  As we will see, our simulations
are not yet good enough to allow the $r$ dependence of
$\alpha$ to be resolved, and so we take it to be a constant.  
The simulations are, however, sufficiently
resolved to see a strong dependence of $f_s(r)$ on $r$, which we
detail below.

The distribution $P(\rho;r)$ in Eq.~(\ref{eqn:pdf}) is
normalized so that $\int_0^\infty \, P(\rho;r)\, d\rho=1$.  It
can be integrated to give the mean density,
\begin{eqnarray}
     \bar\rho(r) &=& \int_0^{\rhomax} \, \rho\,
     P(\rho)\, d\rho \nonumber \\
     &=& f_s \rho_h + \nonumber \\
     & & (1-f_s)\rho_h \begin{cases}
     \frac{1+\alpha}{\alpha} \left[
     1 - \left(\frac{\rhomax}{\rho_h} \right)^{-\alpha} \right];
     & \text{$\alpha\neq0$,} \\
     (1-f_s)\rho_h \ln
     \frac{\rhomax}{\rho_h}; & \text{$\alpha=0$}. \end{cases}
\label{eqn:barrho}
\end{eqnarray}
This thus provides a relation between the mean density $\rho_h(r)$ of
the smooth component and the canonical host halo density $\bar\rho$.
Strictly speaking, this relation requires the maximum density
$\rhomax$ out to which the power-law tail extends.
However, we will see that numerically $1-f_s$ is so small that
$\rho_h(r) \simeq \bar\rho(r)$ is a very good approximation for
practical purposes.

\begin{figure}[htbp]
\includegraphics[width=\columnwidth]{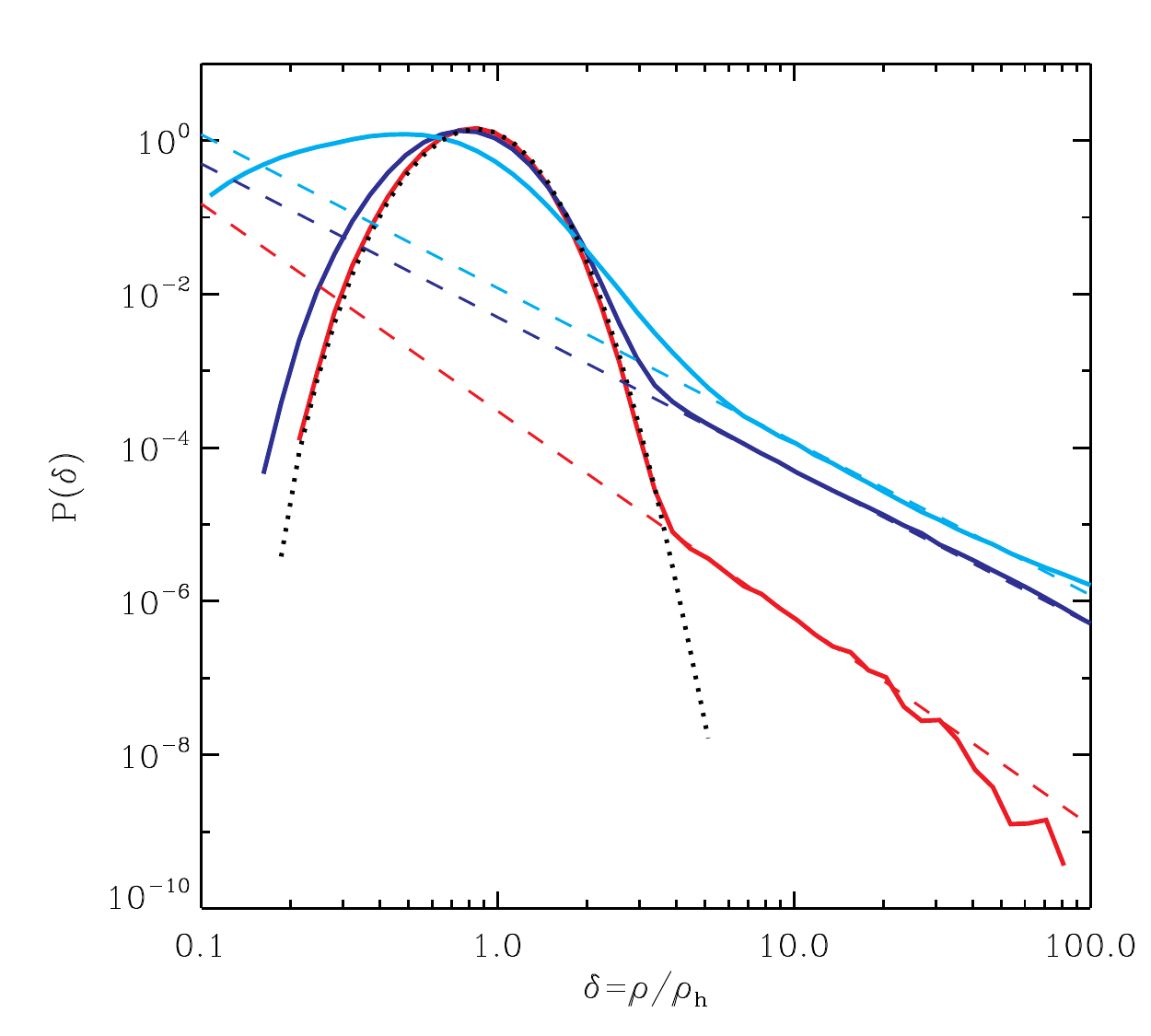}
\caption {\small The probability distribution function $P(\delta)$
  obtained from simulations.  The solid curves are the simulation
  results at $r=300, 100$ and 10~kpc (from top to bottom).  The dashed
  curves show our analytic approximations to the power law tail.  The
  dotted curve indicates the contribution to the finite width of the
  smooth component at 10~kpc from Poisson fluctuations due to the use
  of $N=32$ neighbors in the density estimator.  Note that the mean
  host halo density $\rho_h$ to which the $x$ axis is normalized is
  $\sim175\times$ ($\sim4900\times$) smaller for the 100-kpc (300-kpc)
  curve than in the 10-kpc curve.}
\label{fig:f1}
\end{figure}

\section{Calibration to Simulations}
\label{sec:calibration}

We make use of the Milky Way--like halo in the Via Lactea II
simulation \cite{Diemand:2008in} and calibrate our model to the $P(\rho;r)$
measured from the simulated particle distribution.  Fig.~\ref{fig:f1}
shows the PDF $P(\rho)$ derived from Via Lactea II for three
Galactocentric radii, at 10 kpc (lowest solid line, red), at 100 kpc
(middle line, blue), and at 300 kpc (highest line, cyan).  To obtain
this $P(\rho)$, we consider an ellipsoidal shell that follows an
iso-density contour in the halo.  For each particle in that shell, we
calculate $\delta_i=\rho/\bar\rho$ where $\rho$ is estimated in the
usual manner from the nearest $N$ neighbors using a symmetric SPH
kernel and the median density $\bar\rho$ is obtained from all the
particles in that ellipsoidal shell.  These $\delta_i$ are then binned
in equally-spaced bins in $\log_{10}(\delta)$.  In each of these bins,
we calculate $P(\log_{10}\delta)=\sum_i \delta_i^{-1}$ where the sum
is over all particles in that bin; the $\delta_i^{-1}$ weighting gives
a volume-fraction distribution.  The distribution in $\log_{10}\delta$
is then converted to a distribution in $\delta$ and normalized.

\subsection{Power-Law Tail}

The central features of Fig.~\ref{fig:f1} relevant here are the
high-density power-law tails predicted by
Ref.~\cite{Kamionkowski:2008vw} (and seen already in simulations
\cite{Vogelsberger2009}). The figure shows that the amplitude of
the high-density power-law tail is larger at larger radii.  This can
be attributed largely to the fact that the mean density $\bar\rho$ is
$\sim175$ times lower at 100~kpc than at 10~kpc, and another factor
$\sim30$ times lower at 300~kpc, and so the ratio of the density in
substructures to the mean density is higher at larger radii.

We now use this simulation to calibrate the analytic model at a
variety of radii $r$, from 4 to 300~kpc. At each radius we fit for the
power law parameters $\alpha$ and $f_s$. We find that at radii greater
than $\sim20$ kpc, the smooth-halo fraction is well approximated by
\begin{equation}
\label{eq:fsa}
     1-f_s(r) = 7\times10^{-3} \left(
     \frac{\bar\rho(r)}{\bar\rho(r=100\,{\rm kpc})} \right)^{-0.26}.
\end{equation}
Note that at radii less than $\sim20$ kpc, $1-f_s(r)$ drops faster
than Eq.~(\ref{eq:fsa}); for example, $1 -f_s(10\,{\rm kpc}) = 4
\times 10^{-4} \approx 1.5 \times 10^{-3} \, (\bar\rho(10\,{\rm
  kpc})/\bar\rho(100\,{\rm kpc}))^{-0.26}$. This close to the center,
however, the clumpiness of the simulated halo is likely artifically
suppressed due to finite resolution effects. The best-fit values of
$\alpha$ are $0.0 \pm 0.1$ at all radii greater than 20~kpc. In the
following, we implicitly assume $\alpha=0$ and the radial dependence
in $f_s$ given by Eq.~(\ref{eq:fsa}).

\begin{figure}[htbp]
\includegraphics[width=\columnwidth]{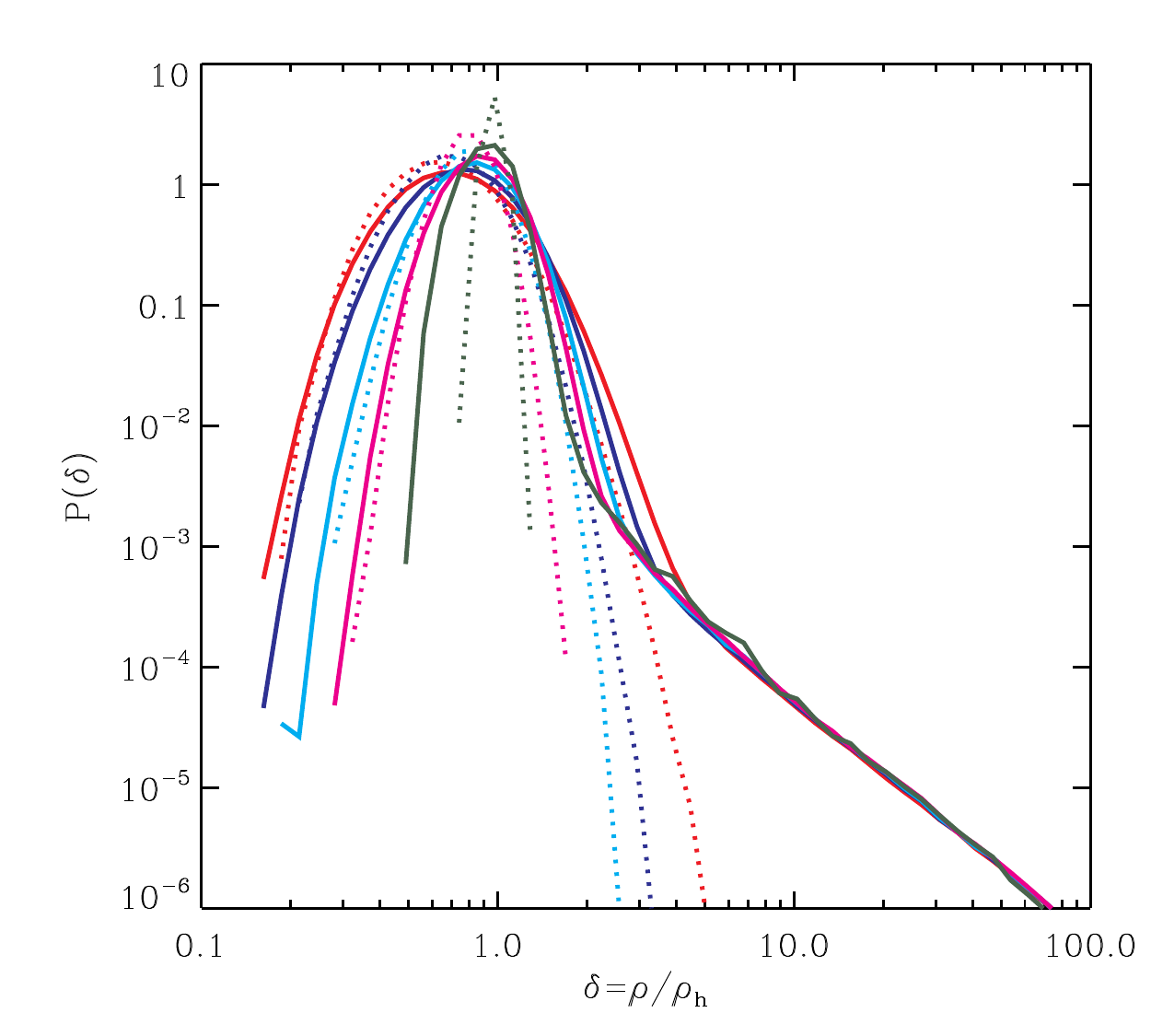}
\caption {\small The probability distribution function $P(\delta)$ at
  100~kpc for particle densities estimated from the nearest $N=(16,
  32, 64, 128, 1024)$ neighbors.}
\label{fig:fnew}
\end{figure}

\subsection{Finite Width of the Smooth Component}

The simulation results shown in Fig.~\ref{fig:f1} show a finite width
$\Delta$ for the smooth component.  However, care must be taken as
Poisson fluctuations due to the finite number $N$ of nearest neighbors
in the density estimator will also contribute to the width. In
Fig.~\ref{fig:fnew} we show $P(\delta)$ at 100~kpc for densities
determined with $N =$ 16, 32, 64, 128, and 1024. The dotted curves
indicated the expected contribution to the width from Poisson
fluctuations (and note that the true and Poisson widths should add in
quadrature), which we obtained by running the density estimator on a
randomly distributed sample of $10^6$ particles. As $N$ is increased,
the width of the smooth component decreases, but not quite as fast as
the Poisson fluctuations, and by $N=1024$ it is clear that the true
width has been resolved to be about $\Delta\simeq0.2$. At 10~kpc (not
shown here) it remains unresolved, and we conclude only that
$\Delta\lesssim0.2$ at radii less than 100~kpc.

\section{Annihilation Boost Factor}
\label{sec:boostfactor}

We now consider the boost of the annihilation rate in a halo
with substructure relative to the rate in the canonical
smooth-halo model.  

The annihilation rate (per unit volume) at any point in the
Galactic halo is
\begin{equation}
     \Gamma = \VEV{\sigma v} \frac{\rho^2}{2 m_\chi^2},
\label{eqn:canonicalGamma}
\end{equation}
where $\VEV{\sigma v}$ is the annihilation cross section (times relative
velocity $v$, averaged over the velocity distribution of the
halo, and $m_\chi$ is the WIMP mass; i.e. $\VEV{\sigma v} =
\int_0^\infty\, dv\, f(v)\, \sigma v$, where $f(v)$ (normalized
to $\int_0^\infty f(v)\, dv =1$) is the WIMP pairwise velocity
distribution at that point in the Galactic halo.  For the
canonical WIMP, $\sigma v$ is approximately velocity independent
at galactic-halo velocities, and so $\VEV{\sigma v}=(\sigma v)$,
a constant.  If $\sigma v$ depends on velocity, then
$\VEV{\sigma v}$ may vary from one point in the halo to another
due to the possible variation of $f(v)$ with position in the
halo.

\subsection{Standard velocity dependence}

Let us first consider the canonical WIMP where the cross-section
factor $\VEV{\sigma v}$ is velocity-independent.  In this case,
the change in the velocity dispersion that may accompany
clumping is irrelevant.  The total annihilation rate in some
volume $V$ is then proportional to a volume integral of the
density squared---i.e., $\propto \int \, \rho^2\, dV$.  The
annihilation rate for a fixed total mass of dark matter enclosed
within a given volume will thus be enhanced by a ``boost
factor'' if the matter is not uniformly distributed within that volume.
For the canonical WIMP, this boost factor $B(r)$ is the ratio 
\begin{equation}
     B(r) = \frac{ \int\, \rho^2\, dV}{\int\, [\bar\rho(r)]^2\, dV}
     =      \int_{0}^{{\rhomax}}
     P(\rho,r)\, \frac{\rho^2}{[\bar\rho(r)]^2}\,d\rho,
\label{eqn:standardBF}
\end{equation}
where the second equality follows since $P(\rho,r) = (1/V)(dV/d\rho)$
(with $V$ the volume in the halo). The quantity $\rhomax$ is the
maximum density. We take that to be $\sim 80$~GeV~cm$^{-3}$, which is
about five times the mean virial density of the rare first collapsed
structures in the Universe (178 times the mean density of the Universe
at $z \approx 40$); we will say more later about
$\rhomax$.

In the standard WIMP model, where $\VEV{\sigma v}$=constant, the
boost factor $B(r)$ is given by (noting that $\bar\rho \simeq
\rho_h$) 
\begin{eqnarray}
      B(r) &=& f_s e^{\Delta^2} \nonumber \\
      & & +
      (1-f_s)\frac{1+\alpha}{1-\alpha} \left[
      \left(\frac{\rhomax}{\rho_h} \right)^{1-\alpha}
      -1 \right].
\label{eqn:rhosquared}
\end{eqnarray}
There is $r$ dependence in this boost factor via Eq.~(\ref{eq:fsa}) in
$f_s$, via Eq.~(\ref{eq:NFW}) in $\rho_h \approx \bar\rho$, and in
principle also in $\Delta$ although it turns out to be negligible.

There are two contributions to this boost factor:  The first
comes from the finite width $\Delta$ of the smooth
component---i.e., $B_s= f_s e^{\Delta^2}$---and it depends
very strongly on $\Delta$.  Given that we find
$\Delta\lesssim0.2$ in our simulations, we infer that the boost
factor due to the finite width of the smooth component is no
more than a few percent.

We now turn to the second term, that due to substructure, the central
focus of this work.  Given that $\rhomax \gg \bar \rho$ in
the Milky Way halo, the boost factor would be essentially independent
of $\rhomax$ if $\alpha$ were $\alpha>1$.  However, we
find in the simulations that $\alpha<1$ in a Milky-Way--like halo, in
which case the boost-factor results for the canonical WIMP do indeed
depend on $\rhomax$.  The fact that the integral in
Eq.~(\ref{eqn:standardBF}) is dominated by the high-density end
implies that the boost factor due to clumping is determined primarily
by the early-collapsing highest-density (and lowest-mass)
substructures.

\begin{figure}[htbp]
\includegraphics[width=\columnwidth]{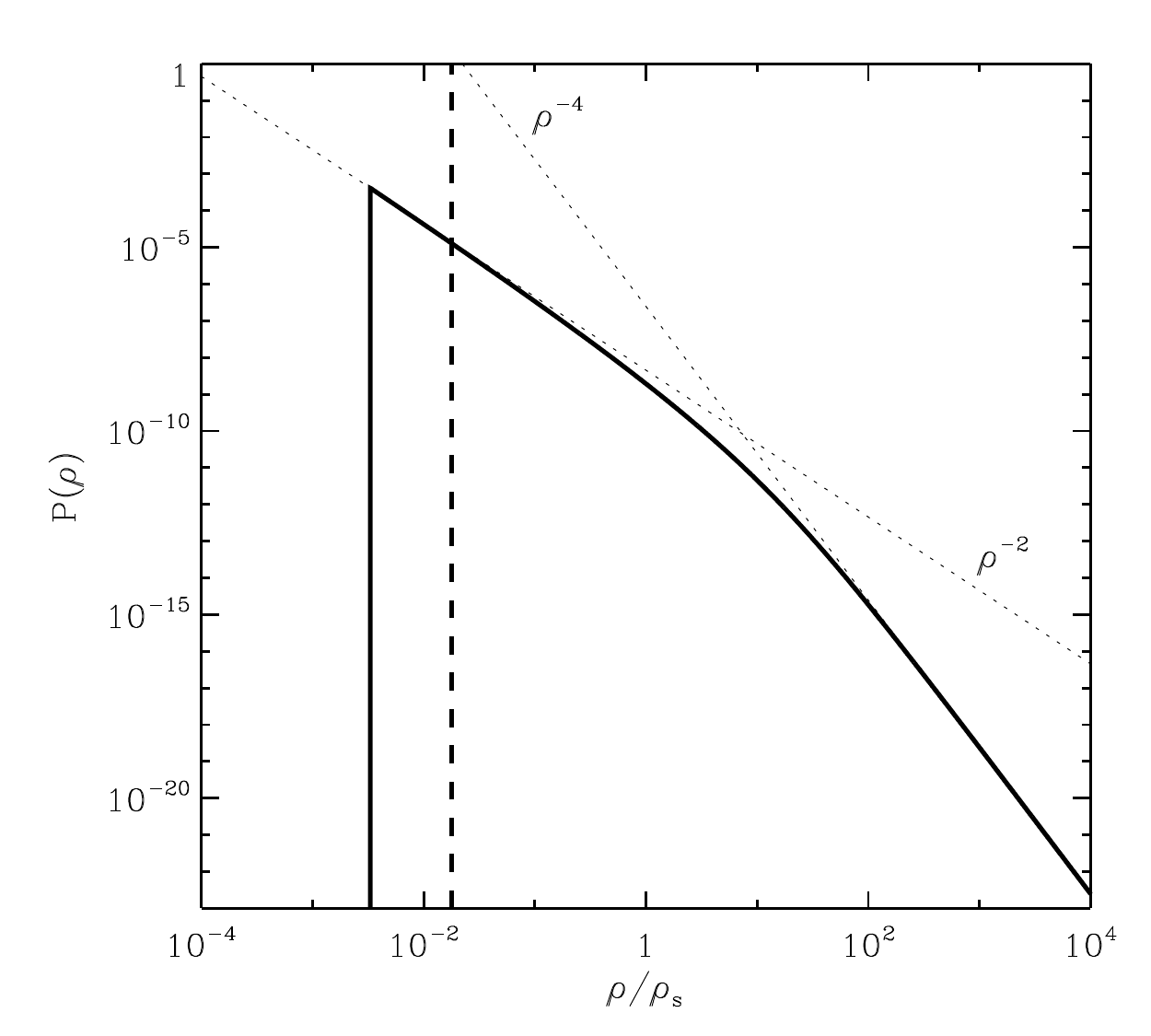}
\caption {\small The density probability distribution function
  $P(\rho)$ for an isolated NFW halo with concentration $c=10$,
  truncated at $r_{\rm vir}$. The $P(\rho) \sim \rho^{-2}$ behavior
  continues beyond $\rho_{\rm vir}$ (vertical dashed line) to $\sim
  \rho(r_s)=1/12\;c^3/f(c)\;\rho_{\rm vir}$.}
\label{fig:Prho_NFW}
\end{figure}

What is an appropriate value for $\rhomax$? A first guess might be the
virial density of the earliest-collapsing halos, $\rho_{\rm vir} = 178
\; \rho_{\rm crit}(z_c) \simeq 16$~GeV~cm$^{-3} \; (z_c/40)^3$ (for a
collapse redshift $z_c \gg 0$). However, depending on its density
profile, most of a halo's \textit{volume} might have considerably
higher densities. In Fig.~\ref{fig:Prho_NFW} we show the density
probability distribution $P(\rho) \equiv 1/V dV/d\rho$ for an isolated
NFW halo. At low densities, in the outskirts of the halo where $\rho
\sim r^{-3}$, the density probability falls as $\rho^{-2}$, matching
the power law tail of the local density probability function
(Eq.~(\ref{eqn:pdf})). In the innermost regions of the halo, where
$\rho \sim r^{-1}$, we have $P(\rho) \sim \rho^{-4}$. The transition
between these two regimes occurs at $\rho_s = 1/12 \; c^3/f(c) \;
\rho_{\rm vir}$, where $c=R_{\rm vir}/r_s$ is the concentration of the
halo and $f(c) = \ln(1+c) - c/(1+c)$. We set $\rhomax = \rho_s = 1/12
\; c^3/f(c) \; \rho_{\rm vir}$, which then depends only on the
concentration with which the earliest-collapsing halos are
born. Numerical simulations
\cite{Diemand:2005vz,*Gao:2005hn,*Diemand:2006ey} indicate low natal
concentrations of $c \approx 2 - 5$, corresponding to $\rhomax \approx
1.5 - 11 \; \rho_{\rm vir}$. For definiteness, we pick an intermediate
value of $c=3.5$ and $z_c=40$, giving $\rhomax = 80$~GeV~cm$^{-3}$.

\begin{figure}[htbp]
\includegraphics[width=\columnwidth]{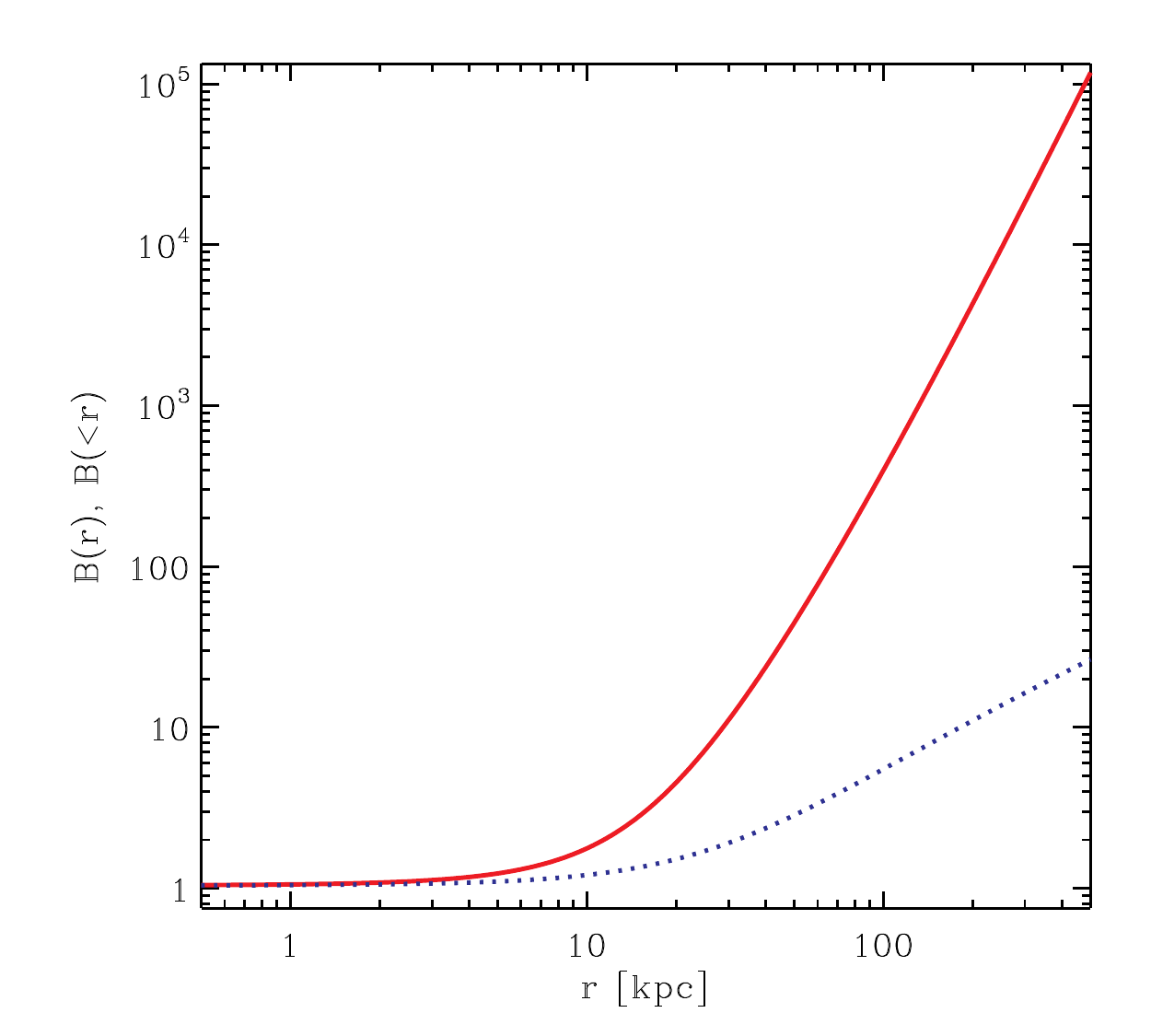}
\caption {\small The local substructure boost $B(r)$ (\textit{solid})
  and the cumulative luminosity boost $B(<\!r)$ (\textit{dotted}), as a
  function of radius.}
\label{fig:BnoS_vs_r}
\end{figure}

Close to the center, the clumped fraction $1-f_s$, as determined from
the numerical simulation, is so small that the boost factor remains
close to unity. As the mean halo density $\rho_h$ decreases with
radius while the clumped fraction increases (cf.  Eq.~(\ref{eq:fsa})),
the local boost factor grows considerably in the outer regions of the
halo. Note, however, that the total luminosity of the halo does not
increase in proportion to this local boost factor. The overall
luminosity is dominated by radii $\lesssim r_s$, and the total boost
from substructure within a radius $R$ must be evaluated numerically,,
\begin{equation}
B(<\!R) = \frac{\int_0^R B(r) \, \rho(r)^2 \, r^2 \, dr}{\int_0^R \rho(r)^2 \, r^2 \, dr}.
\end{equation}
We show in Fig.~\ref{fig:BnoS_vs_r} the differential and cumulative
luminosity boost factor as a function of radius for the Via Lactea II
host halo, assuming $\alpha=0$ and $1-f_s(r)$ as given in
Eq.~(\ref{eq:fsa}). The boost factor remains close to unity in the
center and only reaches 1.5 at the Sun's distance of
8~kpc\footnote{It would be only 1.14 using the value of $1-f_s$
  measured at 10 kpc, instead of Eq.(\ref{eq:fsa}).}, implying that if
the WIMP annihilation cross section has the canonical $\VEV{\sigma v}
\propto $ constant dependence on the velocity, the local boost from
substructure is unlikely to provide the missing factor of 100-1000
needed to explain the cosmic-ray anomalies (see also \cite{Springel:2008zz,*Brun:2009aj}). The total luminosity of the halo,
however, can be appreciably boosted by substructure. The cumulative
boost $B(<r)$ increases to $\sim 17$ at the virial radius.

Before proceeding further, we note that if $\alpha=0$ (and
$\rhomax \gg \bar\rho$), then the boost factor is $B(r) =
f_s+(1-f_s)(\rhomax/\bar \rho)$, an expression that is
easily understood.  The first term is simply the usual annihilation
rate due to the smoothly-distributed dark matter.  The second is that
due to clumping.  If the integrand in Eq.~(\ref{eqn:standardBF}) is
dominated by the high-density end, it implies that most of the
annihilation in the clumped component is taking place in the smallest
and densest subhalos.  If so, then the annihilation rate, per unit
volume, from substructure should be proportional simply to the spatial
density of the subhalos (i.e., how many are there per unit volume),
which itself should be proportional to the ratio of subhalo to host
halo density, $(\rhomax/\bar \rho)$; this is consistent
with Eq.~(\ref{eqn:rhosquared}) if $\alpha=0$.

\subsection{Sommerfeld enhancement}

Suppose now that the annihilation cross section is such that the
thermally-averaged cross section is ${\sigma v} \propto \bar
v^{-\beta}$, where $\bar v$ is the rms relative velocity for
annihilating WIMPs.  There will then be an additional enhancement in
the annihilation rate since lower-mass subhalos will have smaller
velocity dispersions.  We account for this additional effect as
follows:
We first recall that the characteristic density of a first-generation
halo collapsing at $z=40$ with a concentration of $c=3.5$ is $\rho_s =
80$~GeV~cm$^{-3}$, and that the corresponding characteristic velocity
is $v_s \equiv \sqrt{G\,M(< r_s)/r_s} \simeq 1.0 \times
10^{-3}$~km~sec$^{-1}$. We then note that a typical Milky-Way host
halo of mass $2 \times 10^{12} \Msun$ and concentration $c=15$ at
$z=0$ has a corresponding characteristic density and velocity of
$\rho_s=0.076$~GeV~cm$^{-3}$ and $v_s=200$~km~sec$^{-1}$. This thus
suggests a rough scaling $v \propto \rho^{-1.75}$. We emphasize that
this scaling is only meant to very roughly capture the relation
between density and relative velocity of the DM particles. In reality
there likely is no such simple one-to-one relationship between these
two quantities, since regions with similar densities can be bound to
subhalos of very different masses and hence have very different
velocity dispersions. However, as will become clear below, our results
are not very sensitive to the exact value of the power law exponent,
as long as regions of higher density (at a fixed galacto-centric
distance) typically have lower velocity dispersions.

In the following, we consider only the Sommerfeld-enhanced boost
factor from the clumped component and disregard the small contribution
from the finite width of the smooth component. We further
assume that the dark-matter velocity dispersion $\bar v_{\mathrm{MW}}
\simeq 220$~km~sec$^{-1}$ in the Galactic halo does not vary with
Galactocentric radius $r$.  Strictly speaking, this constancy does not
hold for a self-gravitating NFW distribution.  Realistically, though,
the Milky Way disk contributes very significantly to the potential in
the inner Galaxy, and so the dark-matter spatial/velocity distribution
in the inner Galaxy cannot be a pure self-gravitating NFW
distribution.  Our assumption of a constant dark-matter velocity
dispersion is probably closer to the truth than the radial change in
the velocity dispersion implied by an NFW distribution.

With these approximations and assumptions, the boost factor is
\begin{equation} 
B(r) =
     f_s + (1-f_s)(1+\alpha) \int_{\bar\rho}^{\rhomax}
     \, \frac{d\rho}{\bar\rho}\, \left(
     \frac{\rho}{\bar\rho}\right)^{-\alpha} \left( \frac{
     v_{\mathrm{MW}}}{v(\rho)} \right)^\beta,
\label{eqn:SomBF}
\end{equation}
which we then integrate with the relation $v(\rho) =
v_0(\rho/\rhomax )^{-1.75}$ to find
\begin{eqnarray}
     B(r) & = & f_s + (1-f_s) \frac{1+\alpha}{1+1.75\,\beta-\alpha}
     \left( \frac{v_{\mathrm{MW}}}{v_0} \right)^\beta \nonumber \\
     & \times & \left[
     \left(\frac{\rhomax}{\bar\rho}\right)^{1-\alpha} \!\!- 
     \left(\frac{\rhomax}{\bar\rho}\right)^{-1.75\,\beta} \right].
\label{eqn:BwithSommerfeld}
\end{eqnarray}
Again, for the values of $\alpha$ we see in the simulation, the
integrand in Eq.~(\ref{eqn:SomBF}) is dominated by the high end,
increasingly so for $\beta>0$.  Now, even though $1-f_s$ may be small
(the clumped fraction is small), the velocity enhancement
$(v_{\mathrm{MW}}/v_0)^\beta$ may be large, even for values
$\beta\simeq 1$.  Again, if the integrand is dominated by the
high-density tail, it implies that most of the annihilation in the
clumped fraction is occurring in the lowest-mass highest-density
regions.  And if so, then the annihilation rate per unit volume should
again be proportional simply to the ratio of subhalo to host halo
density $(\rhomax/\bar \rho)$, which is again implied in
Eq.~(\ref{eqn:BwithSommerfeld}) if $\alpha=0$.

If the mass of the force carrier particle mediating the annihilation
is non-zero, the Sommerfeld effect saturates at a finite velocity
$\vsat$, when the de-Broglie wavelength of the particle becomes
longer than the range of interaction. With such a saturation the
integral in Eq.(\ref{eqn:SomBF}) is split into two parts: one from
$\bar\rho$ to $\rho_{\rm sat} = \rhomax (v_0/v_{\rm
  sat})^{1/1.75}$ with a velocity-dependent Sommerfeld enhancement term
$(v_{\rm MW}/v(\rho))^\beta$, and a second part from $\rho_{\rm sat}$
to $\rhomax$ with a constant enhancement factor of $(v_{\rm
  MW}/\vsat)^\beta$, 
\begin{eqnarray}
B(r) & = & f_s + (1-f_s)(1+\alpha) \left(\frac{v_{\rm MW}}{\vsat}\right)^\beta \nonumber \\
\times & & \hspace{-0.2in} \left[ \frac{1}{1 + 1.75 \beta - \alpha} \left( \left(\frac{\rho_{\rm sat}}{\bar\rho}\right)^{1-\alpha} - \left(\frac{\rho_{\rm sat}}{\bar\rho}\right)^{-1.75\beta} \right) \right. \nonumber \\
& &  + \left. \frac{1}{1-\alpha} \left( \left(\frac{\rhomax}{\bar\rho}\right)^{1-\alpha} - \left(\frac{\rho_{\rm sat}}{\bar\rho}\right)^{1-\alpha} \right) \right].
\label{eqn:BwithSommerfeld_sat}
\end{eqnarray}

\begin{figure}[htbp]
\includegraphics[width=\columnwidth]{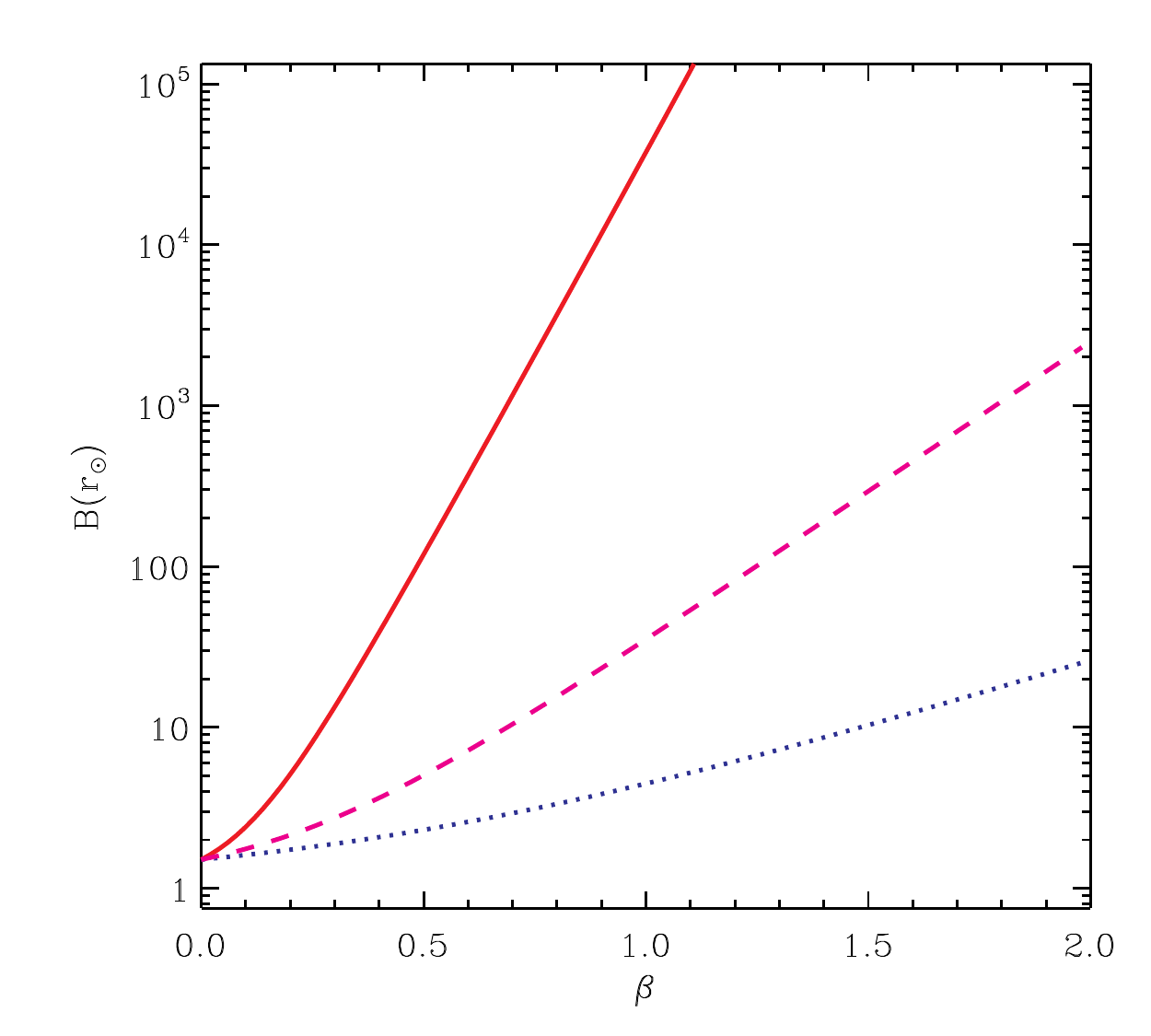}
\caption {\small The boost factor at the solar radius as a function of
  the parameter $\beta$, for no saturation (\textit{solid}), $v_{\rm
    sat}=10^{-4}\,c$ (\textit{dotted}), and $\vsat=10^{-5}\,c$
  (\textit{dashed}). The Sommerfeld-like enhancement grows as
  $v^{-\beta}$ until $\vsat$.}
\label{fig:f2}
\end{figure}

Fig.~\ref{fig:f2} shows the boost factor at the solar radius as a
function of the velocity parameter $\beta$. The solid line depicts the
case without saturation, and the dotted and dashed lines with $\vsat/c
= 10^{-4}$ and $10^{-5}$, respectively. We use
$\rhomax=80$~GeV~cm$^{-3}$ and $v_0=1.0 \times 10^{-3}$~km~sec$^{-1}$
here. With Sommerfeld enhancement it is possible to get very large
substructure boost factors even at the solar radius. \textit{This
  substructure boost, of course, applies in addition to the Sommerfeld
  enhancement of the smooth halo annihilation luminosity.}

\begin{figure}[htbp]
\includegraphics[width=\columnwidth]{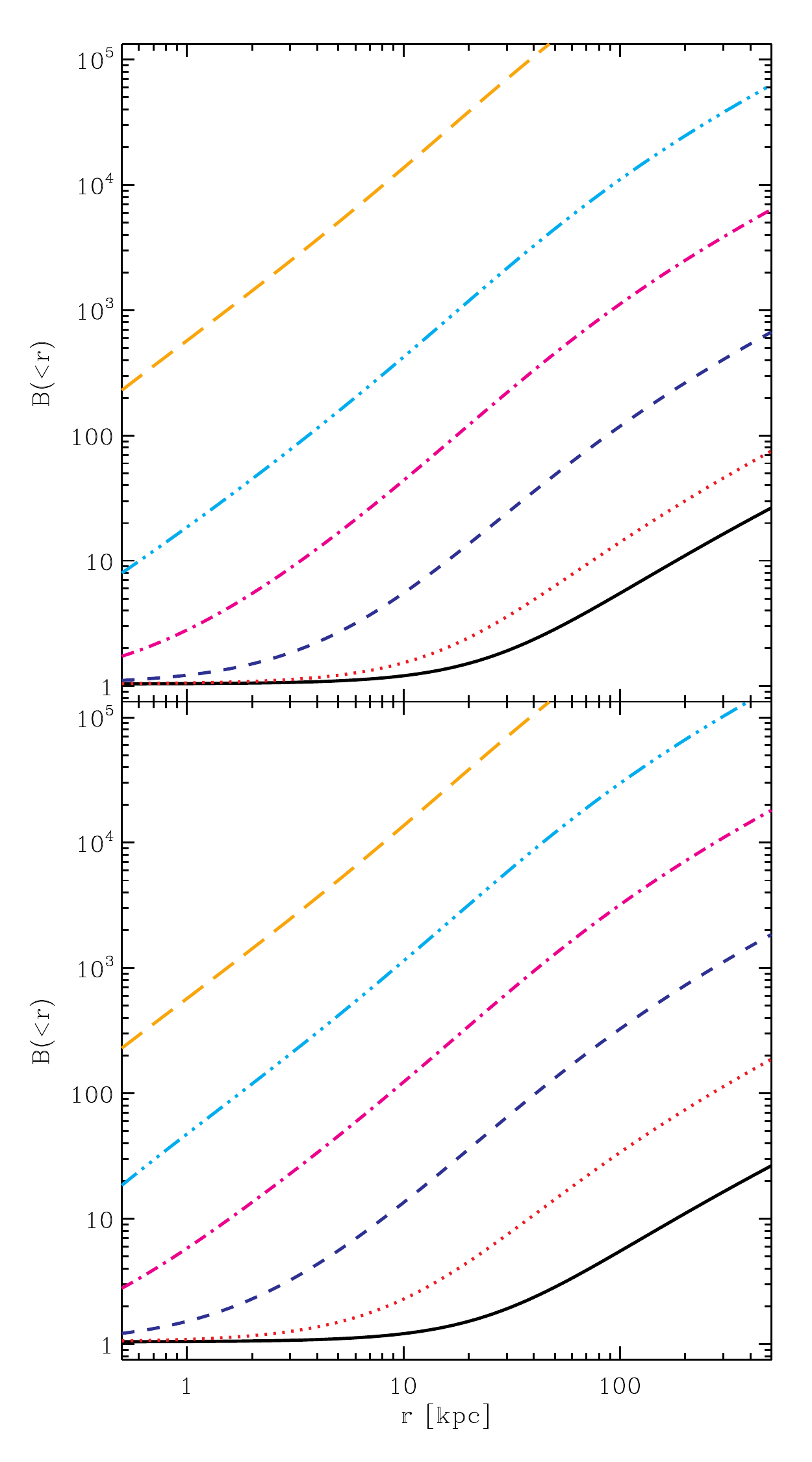}
\caption {\small The cumulative boost factor as a function of distance
  from the Galactic center. In both panels the solid curve represents
  the canonical case ($\beta = 0$). \textit{Top panel:} $B(<\!r)$ for
  different values of $\beta$ with no velocity saturation: $\beta =$
  0.1, 0.3, 0.5, 0.7, and 1.0 from bottom to top. \textit{Bottom
    panel:} $B(<\!r)$ for $\beta=1.0$ and different values of the
  saturation velocity: $\vsat/c = 10^{-4}, 10^{-5}, 10^{-6}, 10^{-7}$,
  and 0 from bottom to top.}
\label{fig:f3new}
\end{figure}

Fig.~\ref{fig:f3new} shows the cumulative boost factor $B(<\!r)$ as a
function of radius. In the top panel we plot curves for $\beta = $ 0,
0.1, 0.3, 0.5, 0.7, and 1.0, assuming no saturation. In the bottom
panel we fix $\beta=1$ (except for the reference $\beta=0$ case) and
vary the saturation velocity, $\vsat/c = 10^{-4}, 10^{-5}, 10^{-6},
10^{-7}$, and 0. Increasing $\beta$, or lowering $v_{\rm sat}$ at a
fixed $\beta$, leads to significant increases in the cumulative boost
factor. For example, the total boost factor from within the virial
radius of $\sim 300$~kpc grows from 17 ($\beta=0$) to $1.3 \times
10^6$ for $\beta=1$ without saturation. Even with $v_{\rm
  sat}/c=10^{-4}$, the $\beta=1$ case still results in about an order of magnitude increase in the total boost, to $\sim 120$.

\section{Application to WIMP models}
\label{sec:wimpmodels}

We now assemble estimates for the numerical values of $v_0$ and
$\rhomax$ for WIMP models.  After freezeout of WIMP annihilation in
the early Universe, the WIMPs may continue to scatter from the more
abundant light standard model particles.  These scatterings suppress
perturbations in the WIMP density on sub-horizon scales until these
elastic-scattering interactions cease; i.e., at kinetic decoupling.
This post-freezeout kinetic coupling of WIMPs suppresses primordial
perturbations on mass scales smaller than $M_c \simeq 33
(T_{\mathrm{kd}}/10\,{\mathrm{MeV}})^{-3}\, M_\oplus$
\cite{Loeb:2005pm}, where $T_{\mathrm{kd}}$ is the kinetic-decoupling
temperature.  The smallest substructures in the Milky Way halo will
therefore have formation masses no smaller than $M_c$. Close to the
halo center, tidal interactions and impulsive stellar encounters may
remove some of fraction of this mass \cite{Zhao:2005mb}, but the dense
cuspy cores contributing the majority of the annihilation luminosity
likely survive \cite{Goerdt:2006hp}. Detailed calculations of the
relevant elastic-scattering reactions shows that in SUSY and UED
models for WIMPs, this mass scale spans the range $10^{-6}\, M_\oplus
\lesssim M_c \lesssim 100\,M_\oplus$ \cite{Profumo:2006bv}, the
precise value depending on the particle-physics details.  Objects in
this mass range undergo gravitational collapse at a redshift $z_c
\simeq 40-\log_{10}(M_c/M_\oplus)$ \cite{Kamionkowski:2008gj}; the
weak dependence of the collapse redshift on $M_c$ arises from the
$n\rightarrow -3$ limit of the primordial power spectrum at small
scales.  These first collapsed objects obtain virial velocities
$v_{\rm vir} = v_0 \simeq 1.0\times 10^{-3}\,(M_c/M_\oplus)^{1/3}
(z_c/40)^{1/2}$ km s$^{-1}$ and virial densities $\rho_{\rm vir}
\simeq 16\,(z_c/40)^3$~GeV~cm$^{-3}$, corresponding to
$\rhomax=80\,(z_c/40)^3\,(c/3.5)^3\,f(3.5)/f(c)$~GeV~cm$^{-3}$.

For a given WIMP model, the cutoff mass $M_c$ can be calculated and
thus $\rhomax$ and $v_0$ obtained.  Given these
parameters, our Eq.~(\ref{eqn:BwithSommerfeld}) can provide the boost
factor, as a function of radius, for given Sommerfeld parameters
$\beta$ and $\vsat$.

For example, suppose we would like a substructure boost of $B(r)\simeq
10^3$ locally to account for reported cosmic-ray anomalies. For an Earth mass cutoff
($M_c=M_\oplus$, $v_0=1.0\times 10^{-3}$ km s$^{-1}$, $\rho_{\rm
  max}=80$~GeV~cm$^{-3}$), taking $\alpha=1$, and a Sommerfeld model
without saturation, the boost factor in
Eq.~(\ref{eqn:BwithSommerfeld}) becomes
\begin{equation}
     B(r) = \frac{(1-f_s)}{1+1.75\beta}
     \frac{\rhomax}{\bar\rho}
     \left(\frac{v_{\mathrm{MW}}}{v_0} \right)^\beta,
\label{eqn:Bapprox}
\end{equation}
which is $\gtrsim 10^3$ for $\beta \gtrsim 0.69$. Likewise, assuming
$\beta=1$, a saturation velocity $\vsat/c \lesssim 3.3 \times 10^{-7}$
is necessary to get a local substructure boost factor greater than
$10^3$.

\section{Angular Dependence of the Gamma-Ray Intensity}
\label{sec:angular}

\begin{figure}[htbp]
\includegraphics[width=\columnwidth]{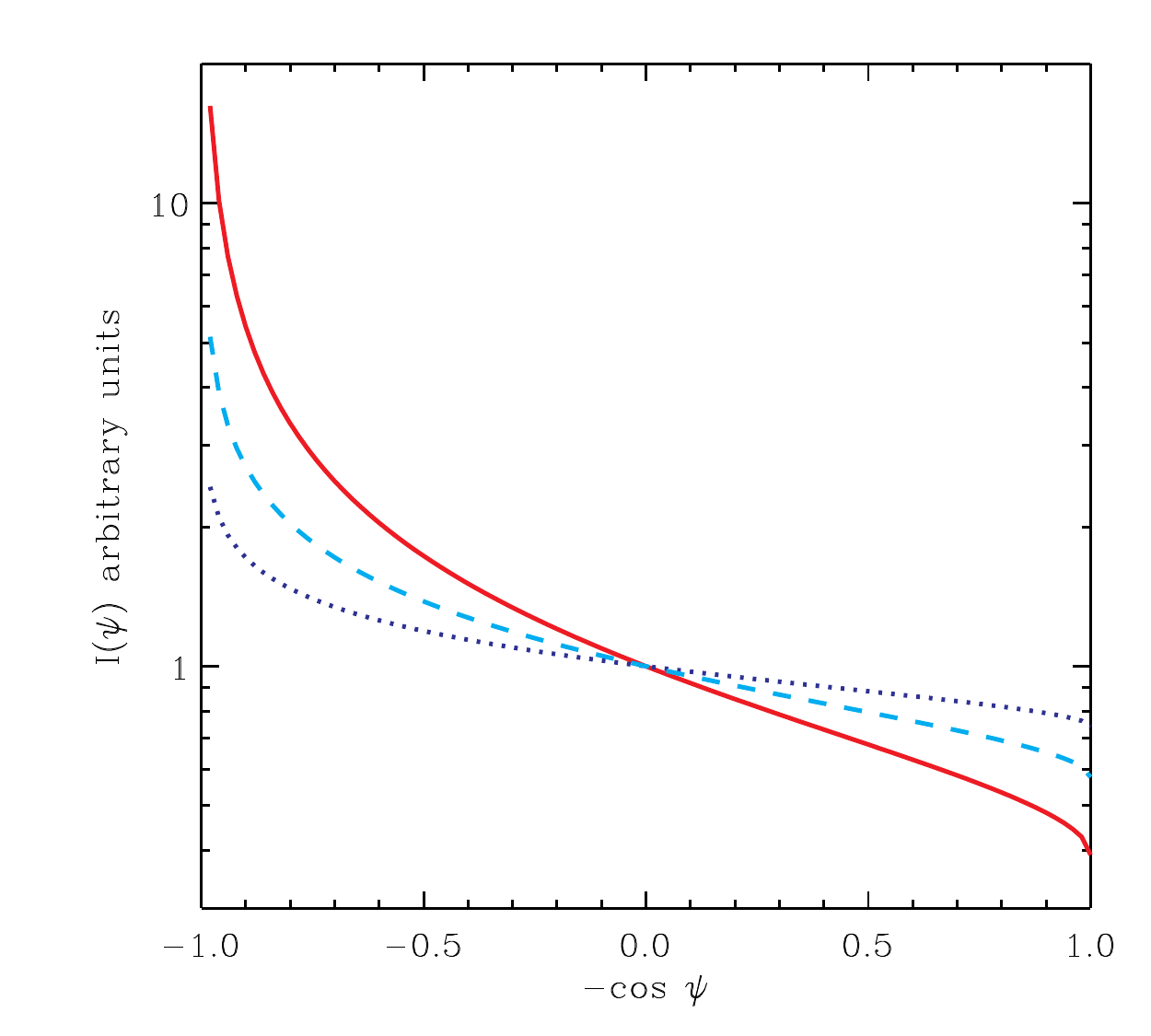}
\caption {\small Intensity of gamma-ray radiation from WIMP
  annihilation in the Milky Way as a function of the (cosine of the)
  angle $\psi$ that the line of sight makes with the Galactic center.
  All three curves are normalized to have the same intensity at
  $\cos\psi=0$.  The top (solid) curve is the intensity due to
  annihilation in a smooth halo; i.e., for $I\propto \int \rho^2\,
  dl$.  The bottom (dotted) curve is that for annihilation entirely in
  subhalos; i.e., $I\propto \int \rho\, dl$.  The middle (dashed)
  curve is for a Sommerfeld-enhanced annihilation [i.e.,
    Eq.~(\protect\ref{eqn:BwithSommerfeld_sat})] with $\beta=1$,
  $\vsat/c=5.0\times10^{-5}$.}
\label{fig:intensity}
\end{figure}

We now consider the dependence of the intensity
(photons cm$^{-2}$~s$^{-1}$~sr$^{-1}$) of gamma-ray radiation from WIMP
annihilation in the Milky Way halo as a function of the angular
separation $\psi$ between a given line of sight and the Galactic
center.  This intensity $I(\psi)$ can be written as an integral
\begin{eqnarray}
     I(\psi) &\propto&  \int_0^\infty \, dl \left[
     \bar\rho\left(\sqrt{l^2+R_0^2 -2 l R_0 \cos\psi}\right)
     \right]^2 \nonumber \\
     & & \times B\left(\sqrt{l^2+R_0^2 -2 l R_0 \cos\psi}\right),
\label{eqn:intensity}
\end{eqnarray}
along a line-of-sight distance $l$.  If the halo is smooth, then
$B(r)=1$, and the integral is $I\propto \int \bar\rho^2 \, dl$,
of the square of the smooth-halo density along the line of
sight.  If all of the annihilation in the halo took place in
highly dense and very small subclumps, then the intensity would
depend on an integral $I \propto \int \bar\rho\, dl$ simply of
the density (rather than density squared). If the intensity
depends on the integral of $\bar\rho^2$, then the intensity will
vary more rapidly with $\psi$, rising rapidly toward the
Galactic center, than if it depends on the integral of
$\bar\rho$, as shown in Fig.~\ref{fig:intensity}.

Most generally, there may be annihilation in both the smooth
component and in the clumped component, in which case the
angular dependence of the intensity will fall somewhere in
between \cite{Lee:2008fm}, as illustrated in
Fig.~\ref{fig:intensity}.  Measurement of this
angular variation may help shed empirical light on the existence
of a boost factor.

\section{Discussion}
\label{sec:discussion}

It has long been recognized that there may be a hierarchy of structure
in the Milky Way halo, with substructure existing all the way down to
the mass scale of a fraction of an Earth mass, more than 22 orders of
magnitude from the $\sim10^{12}\,M_\odot$ Milky Way halo mass.  It has
also been noted that this substructure may have serious implications
for the dark-matter--annihilation rate in the halo.  If the WIMP has
the canonical velocity-independent $\VEV{\sigma v}$, then the boost
factor could be as high as $(\rhomax/\bar \rho)\sim 200$,
if all of the substructure was preserved.  If the WIMP has a
Sommerfeld enhancement to the annihilation rate, the boost factor
could be even higher, and perhaps dramatically so.

Unfortunately, the 22 or more orders of magnitude between the
substructure cutoff mass and the Milky Way mass prevent the
survival of the smallest-scale substructure to be addressed
directly with simulations, and reliable analytic calculations of
the survival fraction are similarly difficult.  In earlier work
\cite{Kamionkowski:2008vw}, we used the nearly self-similar
behavior of hierarchical clustering to develop an analytic
approach to estimate the substructure survival.  The central
result of that work was a prediction that the probability
$P(\rho)$ for a given point in the Galactic halo to have a local
density $\rho$ will have a high-density power-law tail.

Here, we have fit the parameters of that analytic model to new N-body
simulations that can resolve the high-density tail in $P(\rho)$.  The
analytic model then allows us to extrapolate the behavior of the
simulations to mass scales far below the simulation's resolution
scale.  As a result, we have a simple analytic expression for the
distribution of dark-matter densities within the Milky Way halo, as a
function of Galactocentric radius.  The key qualitative result is that
the fraction of the Milky Way volume occupied by substructures is
small ($1-f_s \lesssim 10^{-3}$ in the central regions, $\sim 10^{-2}$
in the outskirts).  In particular, when we calculate the boost factor
for WIMPs with the canonical (i.e., no) dependence of $\VEV{\sigma v}$
on the velocity, we find that it is small: only about 50\% at the
solar radius, and only $\sim 17$ for the total boost within the virial
radius.  The PDF, supplemented with a scaling for the subhalo velocity
dispersion with subhalo density, allows us to also analytically
estimate the boost factor from substructure in the presence of a
Sommerfeld enhancement.  The central result here is given in
Eqs.~(\ref{eqn:BwithSommerfeld},\;\ref{eqn:BwithSommerfeld_sat}), which provide a boost factor in
terms of the power-law index $\beta$ for the velocity scaling of
$\VEV{\sigma v}$, the saturation velocity $\vsat$, the maximum
substructure density $\rhomax\simeq 80$~GeV~cm$^{-3}$, and the
velocity dispersion $v_0$ of the smallest halos.  We estimate
numerically that local (8~kpc) boost factors $\gtrsim10^3$ can be
obtained with $\beta\gtrsim0.69$, or for $\beta=1$ with $\vsat/c < 3.3
\times 10^{-7}$.

We also discussed the finite width of the density distribution
smoothly-distributed component of halo dark matter, but find
that the effects of this finite width on the boost factor are
small.

There are several caveats to our results and several
improvements that can be pursued in future work.  (1)
Hierarchical clustering is not a precisely self-similar
process.  In particular, given that the primordial mass power
spectrum $P(k) \to k^n$ has a power-law index that becomes
smaller at higher $k$ (smaller distance/mass scales, earlier
collapse times), the power-law index $\alpha$ in our work will
probably have some scale dependence, becoming, if anything,
larger at higher densities.  Taking this into account, our
estimates for the boost factors due to substructure are probably
on the high side.
(2) The small-scale mass cutoff $M_c$ for canonical WIMPs
\cite{Profumo:2006bv} may not apply if the new interactions required
for a Sommerfeld enhancement are taken into account.  The cutoff mass
may therefore be very different.  If it is much larger, than the boost
factors will be reduced (substructures will not extend to such small
scales).  If the cutoff mass is much smaller, the boost factor may be
increased relative to our estimates, but, given the weak dependence of
$z_c$ and thus $\rhomax$ on $M_c$, not by much.  (3) While
identification of the high-density power-law tail in $P(\rho)$ and
measurement of its small amplitude is a big step forward, our N-body
measurement of the parameters required to describe the distribution
$P(\rho,r)$ can certainly be improved upon.  It will be important in
future work to measure $f_s(r)$ more precisely, and to measure and
determine the $r$ dependence of $\alpha$ and $\Delta$. (4)
Fig.~\ref{fig:Prho_NFW} indicates that at densities above $\rhomax$,
the $\alpha=0$ power law for $P(\rho)$ will steepen to something
closer to $\alpha=2$. This steep high-density tail is due to the
small-r $\rho \sim 1/r$ dependence in the earliest NFW halos. For a
canonical WIMP with a velocity-independent $\VEV{\sigma v}$, this
steepening will not affect the results, since the integrand in
Eq.~(\ref{eqn:standardBF}) is proportional to $\rho^{-\alpha}$ and
hence dominated by the low density (i.e., near $\rhomax$) end of this
steep tail. With a Sommerfeld enhancement, however, the integrand is
proportional to $\rho^{1.75\,\beta - \alpha}$ (Eq.~(\ref{eqn:SomBF})),
which for $\alpha=2$ is dominated by the high-density end if
$\beta>8/7$. In other words, if $\beta>8/7$, the substructure
annihilation enhancement will be dominated by the $1/r$ cusps in the
earliest NFW subhalos, rather than the $\alpha=0$ part of $P(\rho)$
that we have considered until now. A more detailed calculation of this
effect would depend sensitively on the smallest radius at which the
$1/r$ NFW behavior is valid, and we leave such a study to future work.

\begin{acknowledgments}
This work was initiated during a workshop, attended by all three
authors, hosted by the Caltech/JPL Keck Institute for Space Studies.
M. Kuhlen thanks the Theoretical Astrophysics Center at UC Berkeley
for support and acknowledges the hospitality of KITP at UC Santa
Barbara, where part of this work was completed.  This work was
supported at Caltech by DoE DE-FG03-92-ER40701 and the Gordon and
Betty Moore Foundation, and in part by the National Science Foundation
under Grant No. PHY05-51164

\end{acknowledgments}

\bibliography{kkk}

\end{document}